\documentclass[preprint,12pt]{elsarticle}

\usepackage{graphicx}
\usepackage{amssymb}

\usepackage{lineno}

\journal{Pub. of the Astronomical Society of the Pacific}

\begin{document}

\begin{frontmatter}


\title{A peculiar interacting Be star binary in the Small Magellanic Cloud \tnoteref{label1}}
\tnotetext[label1]{Based on the ESO proposal 098.D-0099(A)}


\author[1]{Ronald E. Mennickent}
\ead{rmennick@udec.cl}
\author[2]{Thomas Rivinius}
\author[3,4,5]{Lydia Cidale} 
\author[6]{Igor Soszy{\'n}sk}
\author[1,7]{J. G. Fern\'andez-Trincado}
\address[1]{Departamento de Astronom\'{\i}a, Casilla 160-C, Universidad de Concepci\'on, Chile}
\address[2]{European Organization for Astronomical Research in the Southern Hemisphere, Casilla 19001, Santiago 19, Chile }
\address[3]{Facultad de Ciencias Astron\'omicas y Geof\'{\i}sicas, Universidad de La Plata, Argentina}
\address[4]{Instituto de Astrof\'{\i}sica de La Plata, CONICET-UNLP, Argentina}
\address[5]{Instituto de F\'{i}sica y Astronom\'{i}a, Facultad de Ciencias, 
Universidad de Valpara\'{\i}so,  Av. Gran Breta\~na 1111, Casilla 5030, 
Valpara\'{\i}so, Chile}
\address[6]{Warsaw University Observatory, Al. Ujazdowskie 4, 00-478 Warszawa, Poland}
\address[7]{Institut Utinam, CNRS UMR6213, Univ. Bourgogne Franche-Comt\'e, OSU THETA, 
       Observatorie de Besan\c{c}on, BP 1615, 25010 Besan\c{c}on Cedex, France }
\begin{abstract}
We find that the emission line object OGLEJ005039.05-725751.4, a member of the cluster OGLE-CL SMC 64, 
exhibits a peculiar light curve pattern repeating with a recurrence time of 141.45 days. The light curve resembles periodic outbursts with a
duty cycle of 20\%. A second long-cycle of 2500 days is also detected in the photometric dataset. 
Two X-SHOOTER spectra obtained at minimum and maximum  reveal a  Be star dominating at minimum light resembling the Classical Be star 48\,Lib. 
The larger H$\alpha$ emission,  the stronger Na\,D absorption and the appearance of emission in the infrared Ca\,II triplet at maximum, might indicate periodic mass transfer in a complex binary system. 
\end{abstract}

\begin{keyword}
stars:binary \sep stars: emission line \sep stars: evolution


\end{keyword}

\end{frontmatter}

\linenumbers

\section{Introduction}
\label{S:1}

The object OGLEJ005039.05-725751.4 ($V$ = 17.211 mag, $\alpha_{2000}$= 00:50:39.1630, $\delta_{2000}$= -72:57:51.239)\footnote{http://simbad.u-strasbg.fr/simbad/} is a  member of the Small Magellanic Cloud, and it was 
classified as a Be star candidate,
based on optical colors and  light curve variability by
Mennickent et al. (2002). Later, and consistently with this classification, H${\alpha}$ emission was
reported in a slitless survey by Martayan, Baade and Fabregat  (2010).
The object is a member of the 
cluster OGLE-CL SMC 64 (Bica \& Dutra 2000), and possibly is the same object  catalogued as 2MASS J00504006-7257492 (SSTISAGEMA J005040.07-725749.3) with coordinates $\alpha_{2000}$= 00:50:40.067, $\delta_{2000}$= -72:57:49.21
and labeled as a possible red giant branch star in the SIMBAD database. 

 The light curve of OGLEJ005039.05-725751.4 shows an extremely rare, strict repeatability of a Be-star outburst like  brightening with a period of 141.45 days. In order to investigate the nature of this object we obtained two spectra at maximum and minimum light. The analysis of these spectra and survey light curve is reported in this paper.  
A finding chart for the star and its surrounding stellar field is shown in Fig.\,1. 

The paper is organized as follow: in Section 2 we introduce the photometric datasets used in our analysis,  details of our spectroscopic observations are given in Section 3, in Section 4 we present our results including the light curve analysis and the study of the spectroscopic data. In Section 5 a discussion is provided along with a possible interpretation for the system and finally our conclusions are given in Section 6.   

\begin{figure}
\scalebox{1}[1]{\includegraphics[angle=0,width=14cm]{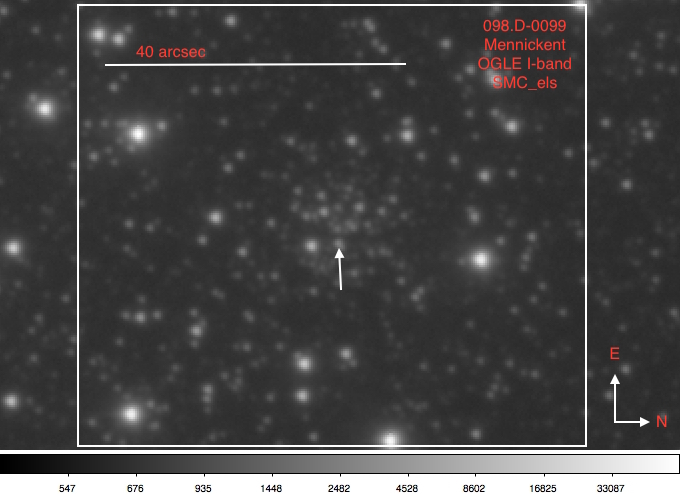}}
\caption{The stellar field around OGLEJ005039.05-725751.4 according to an OGLE $I$-band image. 
OGLEJ005039.05-725751.4 is shown by the arrow at the center of the image. 
   }
  \label{x}
\end{figure}



\section{Photometric data}
\label{S:2}

The photometric time-series data analyzed in this paper were taken from the OGLE project databases.
We included OGLE-II data \citep{2005AcA....55...43S}\footnote{http://ogledb.astrouw.edu.pl/$\sim$ogle/photdb/} and OGLE-III/IV data\footnote{ 
OGLE-III/IV data kindly provided by the OGLE team.}. The OGLE-IV project is described by \citet{2015AcA....65....1U}.
The whole dataset  consists of  1501 $I$-band magnitudes and 143 $V$-band magnitudes taken during a time interval of 17.08 years.
A summary  of these datasets is given in Table\,1.

We considered to study the spectral energy distribution of this object with broad-band photometry provided by the VizieR photometric tool \footnote{http://vizier.u-strasbg.fr/vizier/sed/doc/}. This on-line tool extracts the magnitudes published in surveys and catalogues, based in a search performed around a given location,
considering a searching radius. We find that the scatter shown by the fluxes is quite large, more than expected from the variability of the object, and unfortunately inadequate for our study. It is possible that the presence of several nearby objects in the crowded field is the origin of this scatter, considering that
 automatic photometric algorithms might fail in such circumstances.

 \begin{table}
\centering
 \caption{Summary of survey photometric observations. The number of measurements, starting and ending times for the series and average magnitude and standard deviation (in magnitude) are given. The zero point of HJD is 2\,450\,000.
 Single point uncertainties in the $I$-band and $V$-band are between 4 and 6 mmag.}
 \begin{tabular}{@{}lcrrccc@{}}
 \\ \hline
Database &N &$HJD_{start}$ &$HJD_{end}$&mag &std. &band \\
\hline
OGLE-II  &332 &466.5440 &1871.7550 &16.543&0.126&$I$\\
OGLE-III &732 &2085.9091&4954.8884&16.864&0.174&$I$\\
OGLE-IV &437&5346.9189&6704.5196&16.909&0.143&$I$\\
OGLE-II  &44&466.5830&1543.6290 &  17.037&0.056&$V$\\
OGLE-III &53&3326.5608&4954.8940& 17.053&0.056&$V$\\
OGLE-IV &46&5391.9156&6601.5731& 17.053&0.061&$V$\\
 \hline
\end{tabular}
\end{table}

\begin{table}
\centering
 \caption{Summary of X-SHOOTER spectroscopic observations.The heliocentric Julian day (HJD' $\equiv$ HJD - 2\,457\,900) at mid-exposure  and
wavelength range are given, $R$ is resolving power and $S/N$ signal-to-noise ratio measured around 400 and 680 nm.
 $\Phi$  refers to the phase to the ephemerides given by Eq.\,1.}
 \begin{tabular}{@{}lcccccc@{}}
  \\ \hline
Night &$\Delta \lambda$  & $R$ &exptime  &$S/N$ &HJD' &$\Phi$\\
(2017)&(nm)& &(s)& & & \\
\hline
2/3-Jun &534$-$1020 &5400 & 1260		&35 &07.90506	&0.753\\
2/3-Jun &299$-$556 &7400 & 1161		&55 &07.90454 	&0.753\\
12/13-Jul &534$-$1020 &5400 &1260 		&10 &47.77828 	&0.035\\
12/13-Jul &299$-$556 &7400 & 1161		&50 &47.77776	&0.035\\
 \hline
\end{tabular}
\end{table}


\section{Spectroscopic data}
\label{S:3}

We obtained two spectra for OGLEJ005039.05-725751.4 during the nights of June 2-3 and July 12-13 2017 with the ESO X-SHOOTER spectrograph.
This three-arms echelle spectrograph  is located on Unit Telescope 2 (UT2, Kueyen) of the Very Large Telescope (VLT) at the Paranal Observatory, Chile,
and provides intermediate resolution spectroscopy across a wide wavelength range, from the ultraviolet (UV) to the near-infrared (NIR).
Our observing setup was optimized to get good  recognizing spectra with minimum exposure time 
in the UV and optical ranges neglecting the infrared output and during the minimum (June observations) and maximum (July observations) 
of the photometric cycle described in Section 4. Slit widths of 1.0 arcsec (blue) and 0.9 arcsec (red) were used. 
The spectra were reduced using the X-SHOOTER pipeline, including bias removal, wavelength and flux calibration corrected by atmospheric
differential  refraction. Due to the crowdedness of the field, and to avoid including faint nearby stars in the spectrum, the sky was subtracted only in the visual range. 
One additional step was to remove the barycentric earth's velocity, hence the velocities given here are referred to the  center of mass of the solar system. 
 NIR observations are not considered in this study because of the extremely low signal-to-noise ratio of the spectra.
Our spectroscopic observations are summarized in Table\,2.




\section{Results}
\label{S:4}

 \subsection{Analysis of the light curve}
 
We shifted OGLE-II magnitudes to fit the mean of OGLE-III and OGLE-IV data.
17.5 years of OGLE-$I$ and $V$-band photometry show light modulations similar to outbursts 
recurring with a period of 141.45 days (Fig.\,2). The period was obtained with the  \texttt{PDM}  task \citep{1978ApJ...224..953S}
available in the NOAO software ``Image Reduction and Analysis Facility''
(\texttt{IRAF}\footnote{ IRAF is distributed by the National Optical Astronomy Observatory, which is operated by the Association of Universities for Research in Astronomy (AURA) under cooperative agreement with the National Science Foundation. http://iraf.noao.edu}).  
These ``outbursts" have a duty cycle of 20\% and an amplitude much larger in $I$- than in the $V$-band; their shapes are almost constant, 
with a first excursion to a local maximum, followed by a small brightness decrease and then a second excursion to the maximum
occurring around $I$ = 16.4. The system returns to minimum passing again by a dip and secondary peak 
as revealed in Figs.\,2 and 3. The maxima occur at the same time in both bands, but the
minimum seems to occur earlier in $V$-band than in $I$-band (Fig.\,4).
The $V-I$ color at minimum (+0.12) is compatible with a F1 supergiant but
at maximum the star is  redder; $V-I$ = 0.63 indicates a F5 spectral type (see also Fig.\,2).  In addition to the main light modulation, we find a very long cycle of time scale $T \sim$  2500 days,  more evident in the 
lower envelope of the $I$-band light curve in the top panel of Fig.\,2.  We find the following ephemerides for the maxima:

\begin{equation}
 HJD_{max} = 245\,0587.40 + 141.45\,E  
 \end{equation}

We searched for additional periodicities outside outburst, considering data only in the phase range 0.3-0.9 and removing the 141.45 d periodicity, but 
no additional period was found.

\begin{figure}
\scalebox{1}[1]{\includegraphics[angle=0,width=14cm]{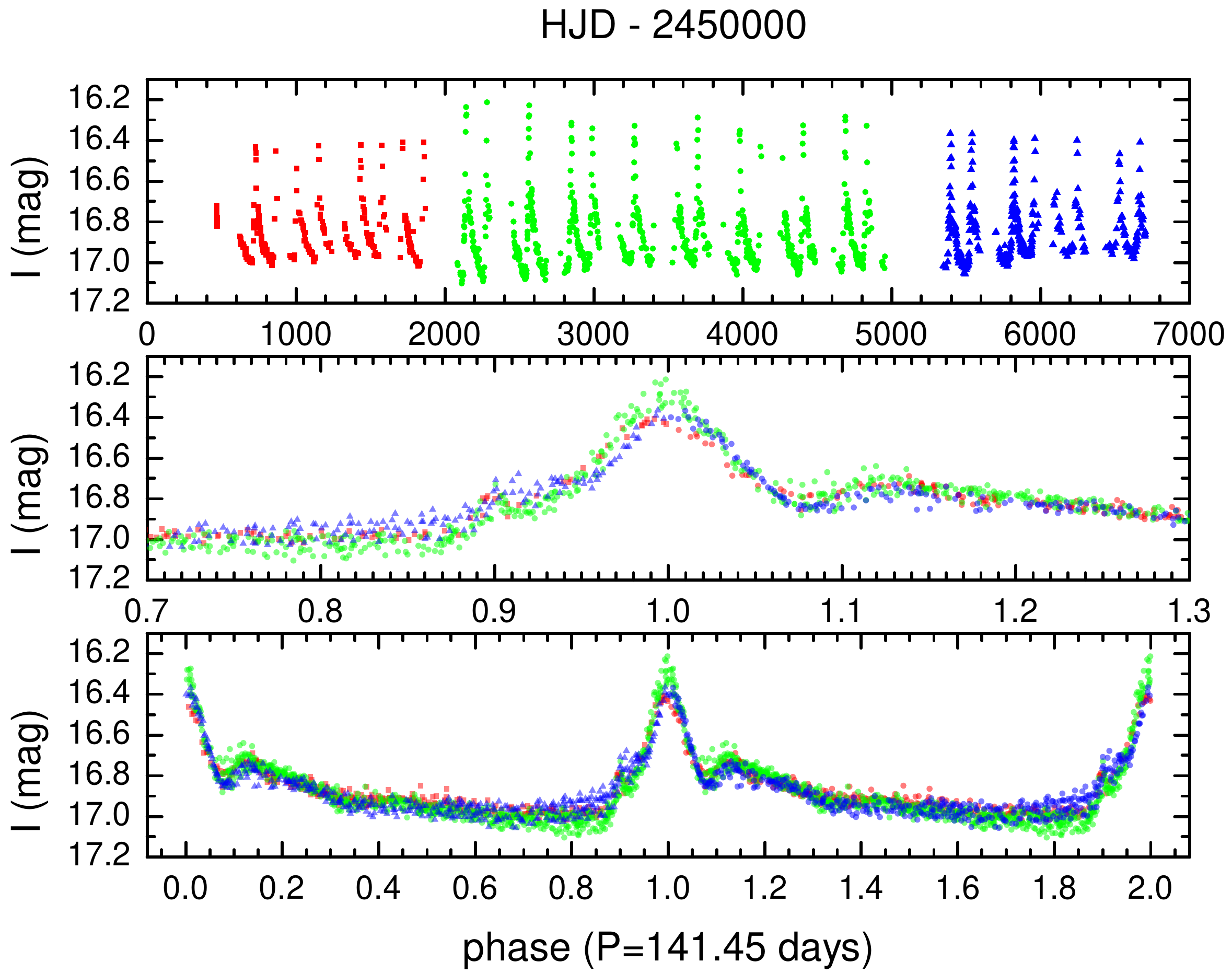}}
\caption{Fig.~1: OGLE $I$-band light curve (top panel), phased with the particular period 
(141.45 days: middle and bottom panels). Colors red, green and blue indicate magnitudes from photometric databases OGLE\,II, III and IV, respectively.
}
  \label{x}
\end{figure}

\begin{figure}
\scalebox{1}[1]{\includegraphics[angle=0,width=14cm]{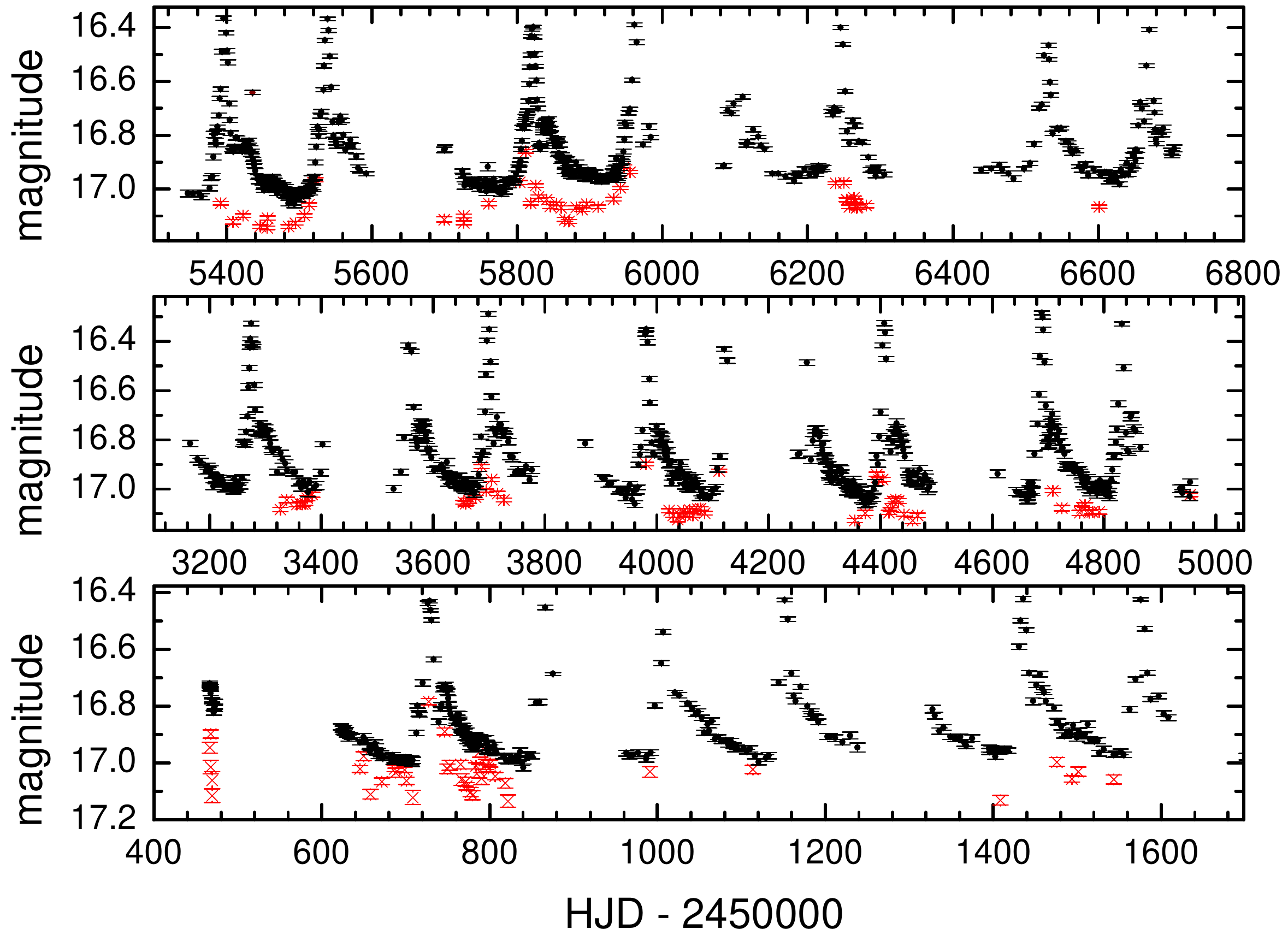}}
\caption{OGLE $I$-band (black dots) and $V$-band (red crosses) light curves at different epochs. Note the smaller amplitude variability in $V$ band. 
}
  \label{x}
\end{figure}

\begin{figure}
\scalebox{1}[1]{\includegraphics[angle=0,width=14cm]{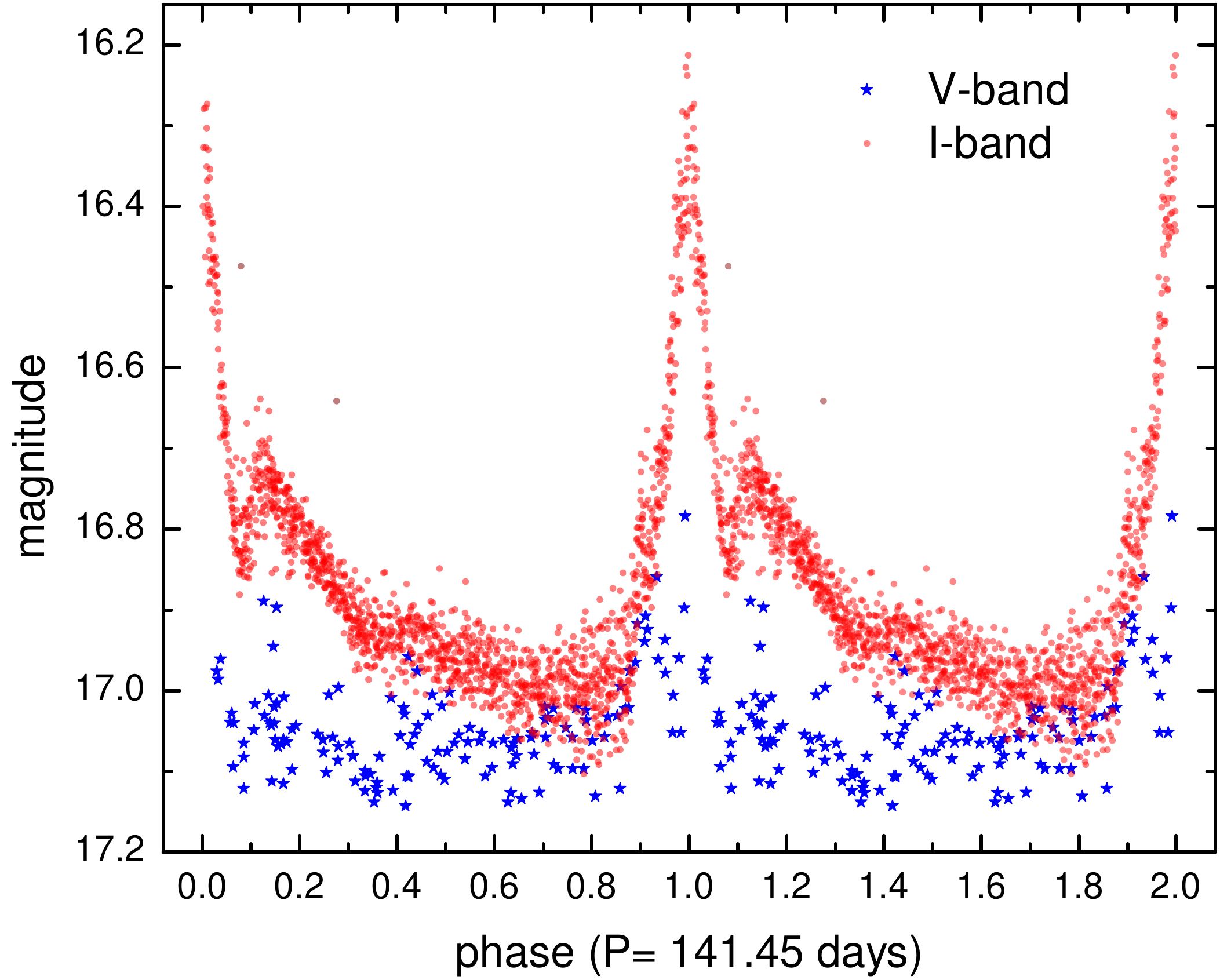}}
\caption{Comparison of $I$-band and $V$-band light curves during the 141.45 days cycle. 
}
%
  \label{x}
\end{figure}

\subsection{Analysis of spectroscopic data}

\subsubsection{The spectrum at minimum}

The spectrum taken at minimum shows emission in H$\alpha$ and He\,I\,{$\lambda \lambda$ 5875}\AA\ 
and He\,I\,{$\lambda \lambda$ 6678}\AA\  
in absorption, revealing an early type, possibly B-type stellar component (Fig.\,5). The Paschen series is seen in absorption, along with double emission 
showing the violet peak larger than the red peak, i.e. $V>R$, where $V$ and $R$ refers to the respective peak intensities.  
We also find O\,I\,{$\lambda \lambda$ 8446}\AA\ as single emission and some metallic double emissions. The spectrum shows a second Balmer discontinuity. This fact, along with the presence of H\,I emission and
sharp absorptions of elements Cr\,II, Ti\,I, Ti\,II, Fe\,I and Fe\,II
reveals the presence of a circumstellar envelope.

 The stellar parameters of the stars were obtained from direct measurements of the Balmer discontinuity. To this aim we 
used the BCD method \citep{1939ApJ....90..627B, 1973A&A....23...69C} that has 
the advantage of studying the Balmer jump to recognize B-type stars with circumstellar envelopes due to the presence of a second component of the Balmer discontinuity 
\citep{1979RA......9..247D, 1991A&A...245..150Z, 2015A&A...577A..45A, 2018A&A...610A..30A}.
A strong second Balmer discontinuity is present in the spectrum taken at the minimum intensity of the light curve ($m_v= 17.2$). 
Based on the height of the Balmer jump $D_{\star}$= 0.17 and its spectral position $\lambda_1$ = 41 \AA, we determined a spectral type B2/3 and temperature $T_{eff}$= 19\,000 K using the recent BCD calibration by \citet{2018A&A...609A.108S}. A spectral type B2/3\,IV
is  consistent with $V$= 17.2 in the SMC, as reveal the study of the star SMC\_SC4 22859 with  $V$= 17.1 \citep{2018MNRAS.476.3555R}.
These authors found 8 solar masses in the middle of the MS for SMC\_SC4 22859, and from our BCD analysis we get about 6.3 masses for our object, which is a
reasonable discrepancy.

We also notice the similarity of the spectrum at minimum with 
the Galactic Be shell-star 48\,Lib (HD\,142983), classified as
B8\,Ia/Iab in SIMBAD. The comparison with the metallic absorption lines of  48\,Lib produced in its envelope
is specially interesting; the comparison spectrum is taken from the UVES-POP catalogue\footnote{http://www.eso.org/sci/observing/tools/uvespop.html}  (Fig.\,6).
 In the case of 48\,Lib, this late classification is purely due to the shell, the actual star has a temperature corresponding to about B3 \citep{2016ApJ...826...81S}. 
The supergiant classification is also only because of the shell, because it has a very strong V/R cycle (relative intensity between emission peaks) that can look like a P Cyg wind profile at times 
\citep{2016ApJ...826...81S}. Some Fe\,II double emission lines with $V > R$ are shown in Fig.\,7 for comparison with the 48\,Lib case. 

\begin{figure*}
\scalebox{1}[1]{\includegraphics[angle=0,width=14cm]{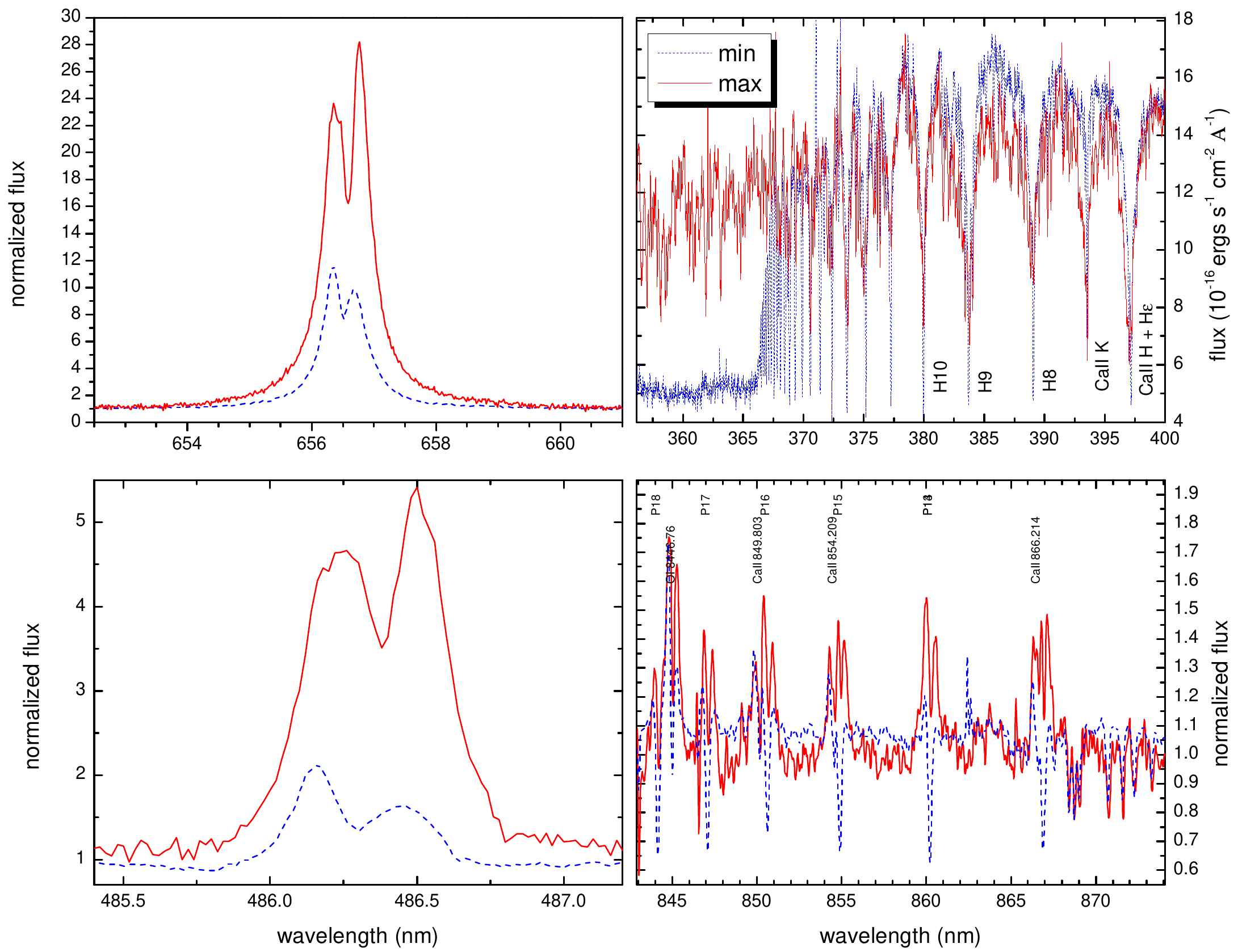}}
\caption{Comparison of spectral regions at maximum (solid red line) and minimum (dashed blue line). The flux calibration of the spectrum at maximum should not be trusted because of the contamination discussed in the text.}
  \label{x}
\end{figure*}

\begin{figure}
\scalebox{1}[1]{\includegraphics[angle=0,width=14cm]{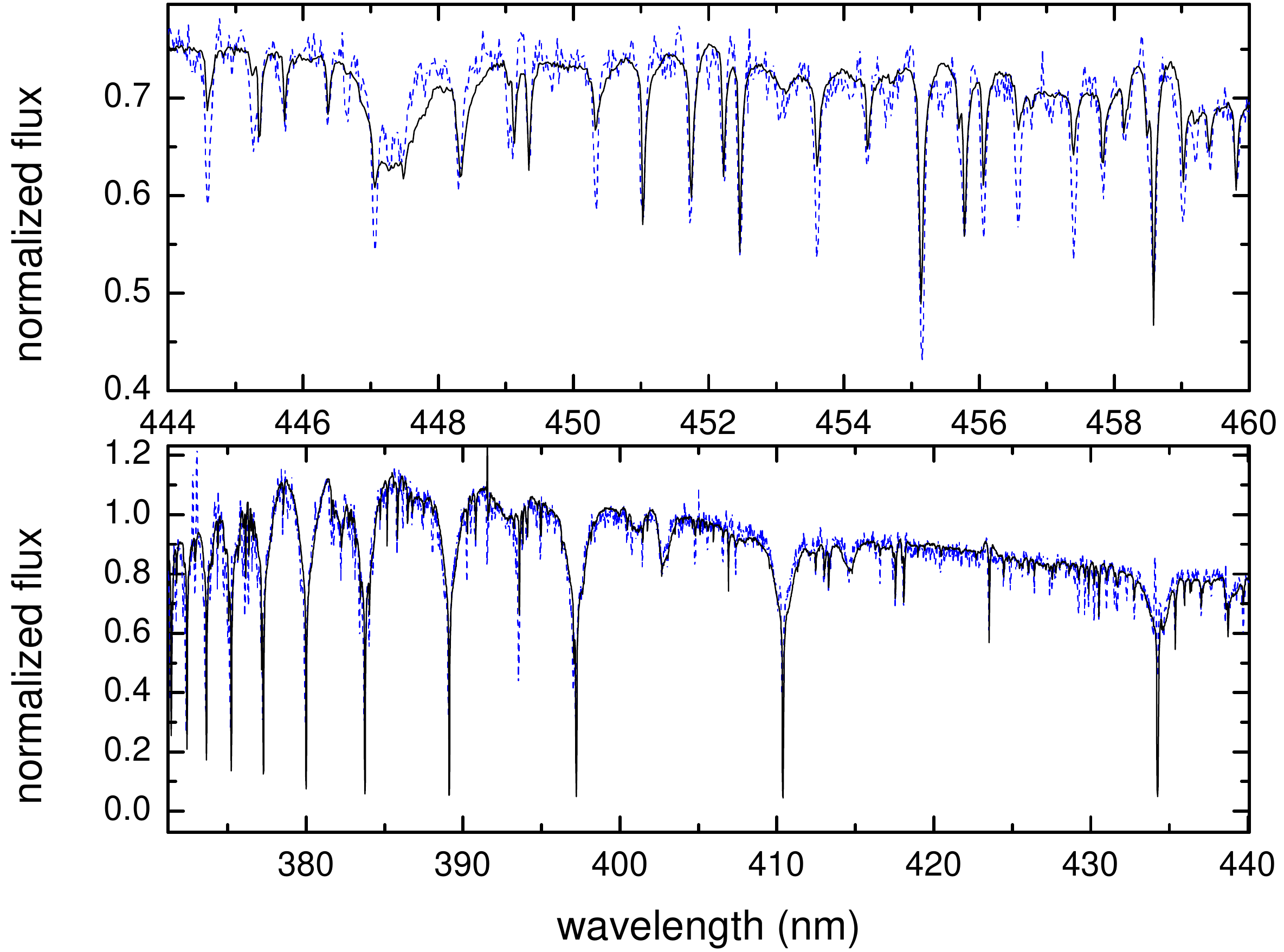}}
\caption{Comparison of the spectrum at minimum (dashed blue line) with the spectrum of 48 Lib (HD\,142983) which is classified B8\,Ia/Iab.}
  \label{x}
\end{figure}

\begin{figure}
\scalebox{1}[1]{\includegraphics[angle=0,width=14cm]{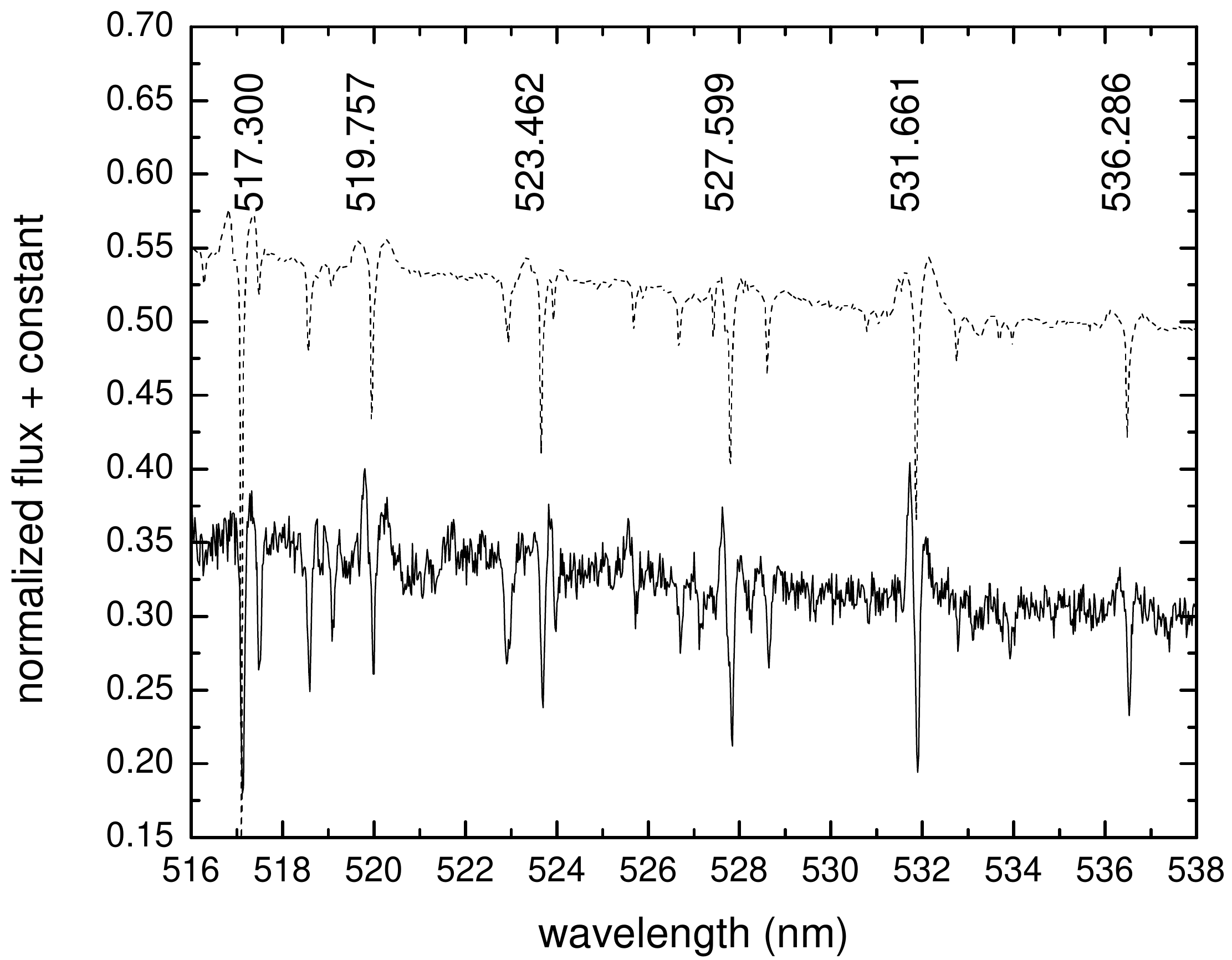}}
\caption{Some Fe\,II double emission lines mostly showing the asymmetry $V > R$ (solid-line) and the comparison with the 48\,Lib spectrum (dashed-line).}
  \label{x}
\end{figure}

\subsubsection{The spectrum at maximum}

We observe much weaker Balmer absorptions than in minimum and 
He\,I\,{$\lambda \lambda$5875} \AA\, 
 appears as double emission
with peak separation of 240 km s$^{-1}$ (Fig.\,8). 
As the spectrum is quite noisy in that region, and other helium lines as He\,I\,{$\lambda \lambda$4471} \AA\, do not show emission, this detection should be considered as tentative only.
Interestingly, as shown in Fig.\,5, double emission is detected with confidence in the infrared Calcium triplet.
At this stage the 
Balmer/Paschen double emissions are in general larger and without the deep absorption cores observed during minimum. The circumstellar metallic absorption lines
still are visible during maximum. 
An additional set of metallic lines appears, which are not observed at minimum, this is specially evident in the blue spectral region. 
Since they span the whole spectrum in the spatial direction at both sides of the stellar profile, this background metallic line spectrum probably arises from reflected moon light. In spite of this contamination, a set of lines characterized by radial velocities displaced by about +140 km/s from the background spectrum are present. Their velocities are  compatible with an origin in the SMC and we assumed they are formed in the system under study.

At maximum  we also find 
a stronger stellar NaD doublet (Fig.\,8). The system NaD lines are clearly distinguished from the sharp Galactic
interstellar NaD lines, that are more clearly visible at minimum. The SMC interstellar components are expected much weaker than the
Galactic ones and are obviously masked by the system components, which are much stronger than expect for a B-type star, and 
are probably formed in the circumstellar medium. 
The increasing
strength at maximum, along with the infrared Ca\,II triplet emission suggest an accretion phenomenon and a formation in a mass stream  as we will discuss in Section 5. 

\subsubsection{General spectroscopic analysis}

Average properties for Balmer and Paschen emission lines are presented in Table\,3, including the ratio between the violet and red peak intensity relative to the normalized continuum $V/R$ $\equiv$ $(I_V-1)/(I_R-1)$. The peak separation increases with the Balmer series order. This gradient, and the existence of double emission, are  typical signatures of a Keplerian Be star disk.  Equivalent widths ($EW$) were measured between the adjacent continuum of the H$\alpha$ line and tracing a line at the base of the other emission lines. 

In general we notice  larger H\,I emission at maximum, along with changes in H\,I line shapes.
At minimum $V$ $>$ $R$ and at maximum  $V$ $<$ $R$ in Balmer lines. However, in Paschen lines  $V$ $>$ $R$ 
in both epochs.  At maximum, we also observe Ca\,II lines as double emissions. While the H$\alpha$ peak separation increases at maximum the opposite is
observed in H$\beta$. H$\beta$ remains almost of the same strength relative to the continuum at both epochs.

Radial velocities for He\,I, Si\,I  and NaD absorption lines are given in Table 4, for 
Fe\,II  emission lines (along with peak separation) are given in Table 5 and for H\,I and metallic absorption lines are given in Table 6.
At maximum we observe a bimodal distribution of the radial velocities suggesting two components, and at minimum only 
one component is clearly observed (Fig. 9); this fact suggests furthermore that the
system is a binary star.

\begin{table}
\centering
 \caption{Average measurements for emission lines and their standard deviations. The methods of barycenter and central minimum
 are indicated for radial velocities with typical error $\pm$ 2 km/s.
 }
 \begin{tabular}{@{}lccccc@{}}
 \\ \hline 
Line &$EW$ &$\Delta \lambda_p$  &$V/R$ &RV(bar) &RV(cen)\\
 & (\AA) & (km s$^{-1}$) &or note &(km s$^{-1}$) &(km s$^{-1}$)  \\
\hline
min  & & & & & \\
H$\alpha$  &-102.8 $\pm$ 0.5 & 153 $\pm$ 3 & 1.19 $\pm$ 0.01  &103 &107\\
H$\beta$  &-5.0 $\pm$ 0.1 & 186 $\pm$ 6 & 1.76 $\pm$ 0.01 &82 &101\\
H$\gamma$ &NA & 225 $\pm$ 2 & $V > R$ &90 &120\\ 
P14 &0.50 $\pm$ 0.05 &- &- &- &132 \\
P17 &-&- &- &- &132 \\
\hline 
max  & & & & & \\
H$\alpha$  &-269.7 $\pm$ 3.0 & 179 $\pm$ 2 & 0.83 $\pm$ 0.04 & 142&139\\
H$\beta$ &-24.0 $\pm$ 0.2 & 159 $\pm$ 1 & 0.83 $\pm$ 0.02 &158 &156\\
H$\gamma$ &-3.9   $\pm$ 0.2 & 188 $\pm$ 5 &V $\sim$ R    &164 &154\\
P14 &-4.7 $\pm$ 0.5  &198 $\pm$ 5 &1.33 $\pm$ 0.01 &135 &148\\
P17 &-5.0 $\pm$ 0.5                              &205 $\pm$ 5 &1.20 $\pm$ 0.01 &144 &149\\
He\,I\,5875 &-1.0 $\pm$ 0.5 &240 $\pm$ 5 &V $\sim$ R &219 &180  \\
\hline
\end{tabular}
\end{table}

\begin{table}
\centering
 \caption{Radial velocities of some absorption lines at maximum and minimum. }
 \begin{tabular}{@{}lr@{}}
 \\ \hline
Line &  RV\\ 
& (km s$^{-1}$) \\
\hline
min & \\
He\,I\,5875 &109 $\pm$ 15* \\
Si\,I\,6347.1  &123.8 $\pm$ 0.2 \\
He\,I\,6678  &127 $\pm$ 3 \\
Na\,I\,5889.95  & 122.7 $\pm$ 0.2 \\
Na\,I\,5895.92 &125.2 $\pm$ 0.2 \\
\hline
Mean (min) & 125 $\pm$ 2 \\
excluding * & \\
\hline 
max & \\
Na\,I\,5889.95 &145.8 $\pm$ 0.2 \\
Na\,I\,5895.92  &147.7 $\pm$ 0.2 \\
Si\,I\,6347.1  &147.0 $\pm$ 0.2 \\
\hline
Mean (max) & 147 $\pm$ 1  \\
\hline
\end{tabular}
\end{table}

\begin{table}
\centering
 \caption{Radial velocities and peak separation for double-emission Fe\,II lines observed during minimum. }
 \begin{tabular}{@{}lrrrr@{}}
 \\ \hline
$\lambda$ (lab) & abs &blue-em &red-em &$\Delta\lambda_p$ \\ 
(\AA)& (km s$^{-1}$) & (km s$^{-1}$)& (km s$^{-1}$)& (km s$^{-1}$)\\
\hline
4233.167&        118.9&	16.2	&222.3 &	206.1 \\
4583.829&	108.9&	-0.4	&241.0 &	241.3\\
4629.336&	116.9&	20.4	&234.8 &	214.4\\
4924.043&	110.1&	4.1	&-- &--\\
5018.434&	116.7&	11.0	&213.5 &	202.5\\
5173.002&	103.7&	-8.1	&-- &--\\
5197.569&	123.6&	10.5	&257.4 &	246.9\\
5275.994&	123.0&	-0.9	&-- &--\\
5316.609/777&	118.6&	17.1	&209.4 & 192.3\\
5362.864&	119.1&	4.5	&-- &--\\
\hline
Average &115.9 &7.5 &229.7&217.2  \\
std &6.4 &9.2 &18.2 & 22.0 \\
\hline
\end{tabular}
\end{table}

\begin{figure}
\scalebox{1}[1]{\includegraphics[angle=0,width=14cm]{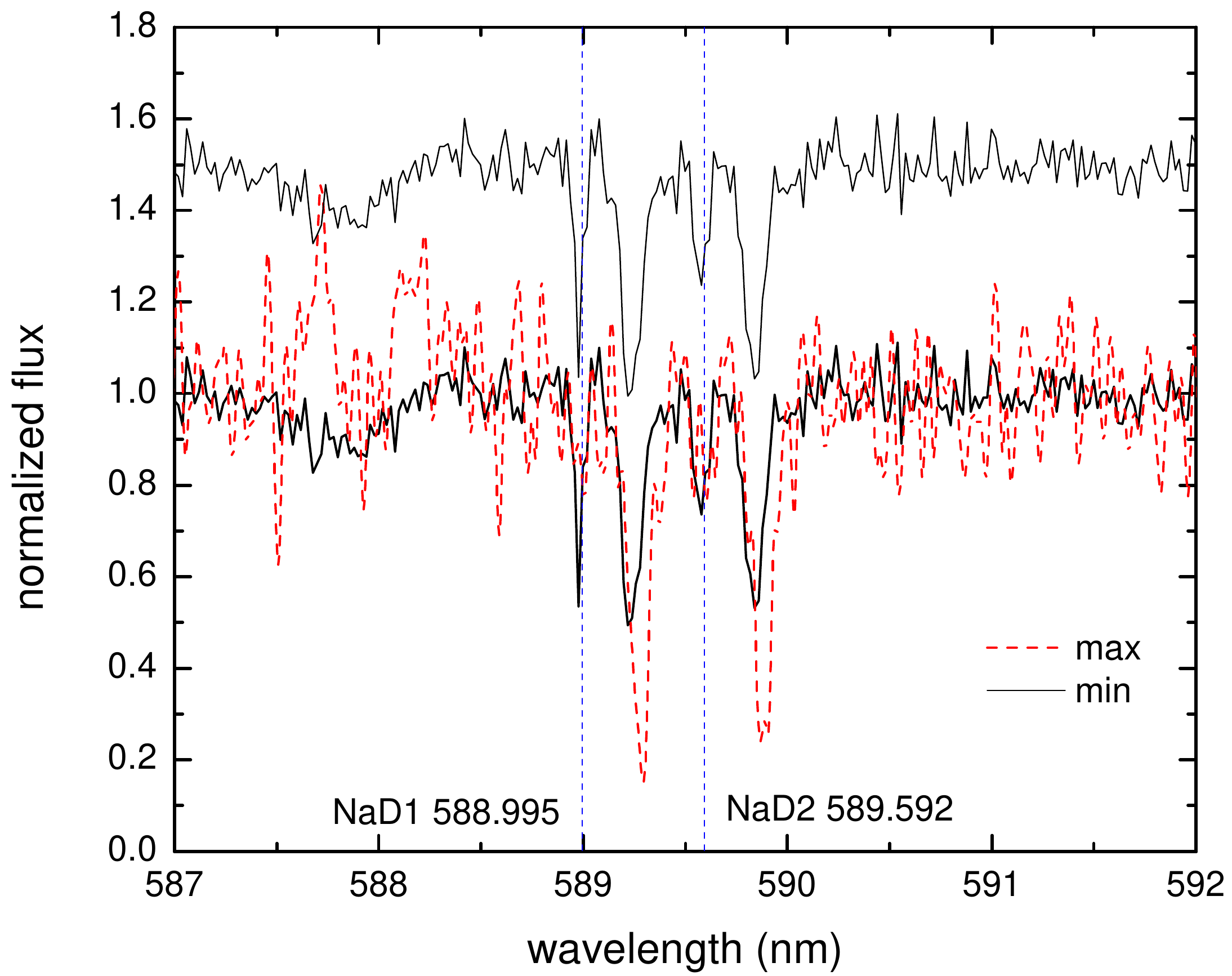}}
\caption{NaD and He\,I\,5875 lines at maximum (dashed red line) and minimum (black). Vertical dashed lines indicate rest wavelengths for D1 and D2 lines of Na. The absorption lines at the rest
wavelengths are due to Galactic interstellar absorption. The spectrum at minimum is plotted twice and shifted for easier comparison. }
  \label{x}
\end{figure}

\begin{figure}
\scalebox{1}[1]{\includegraphics[angle=0,width=14cm]{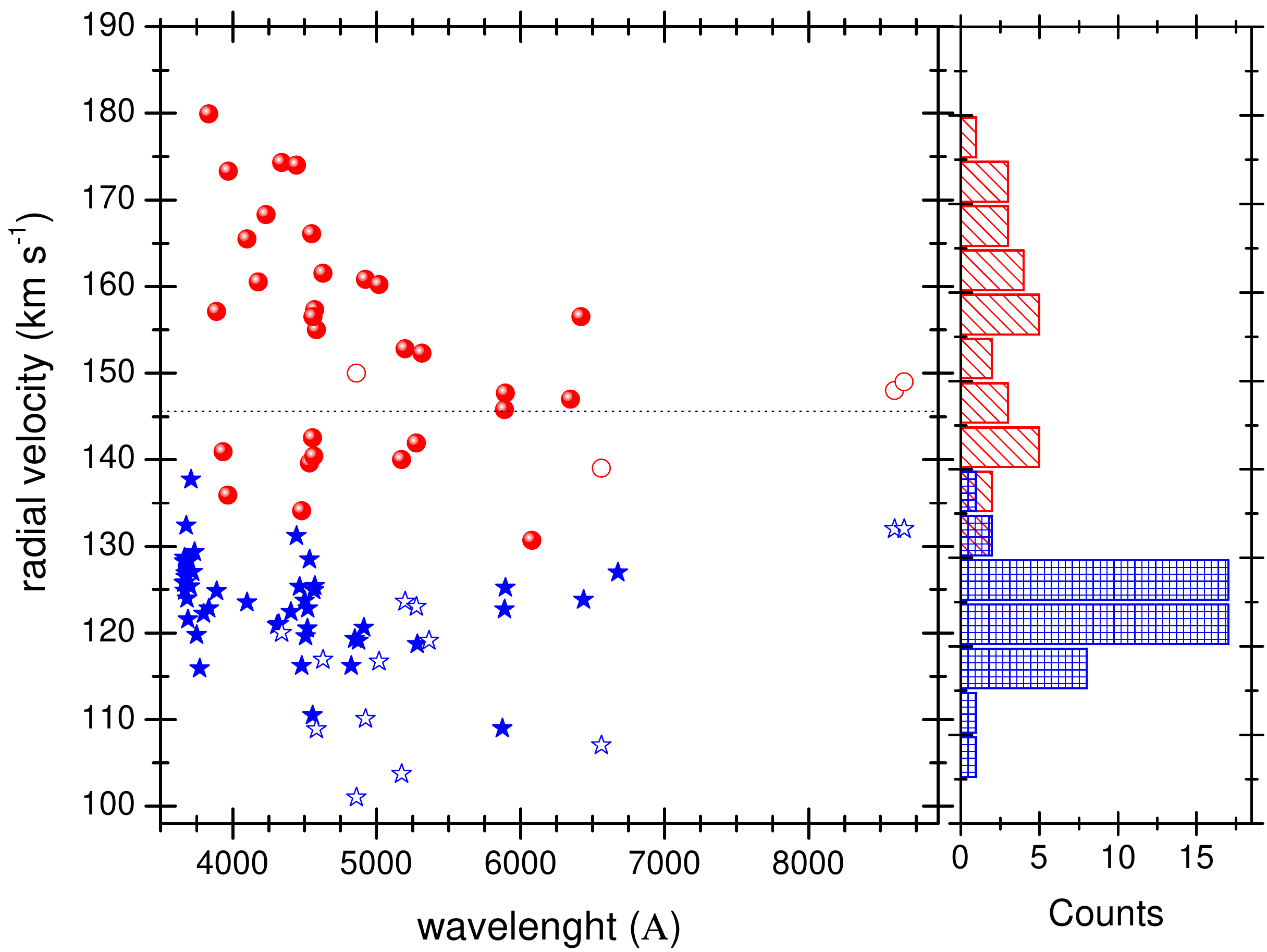}}
\caption{Radial velocities of absorption lines at maximum (red spheres) and minimum (blue stars) and the respective histograms. Open symbols indicating 
central absorption radial velocity of double emission lines are also shown.  The heliocentric velocity of the SMC, 145.6 km s$^{-1}$ \citep{2012AJ....144....4M} is indicated as a dotted line.
}
  \label{x}
\end{figure}

\section{Discussion}
\label{S:5}

At minimum the B-type component dominates the spectrum and the set of radial velocities of the envelope
roughly matches those helium-line velocities associated to the star (Table 4 and 6). This suggests that a circumstellar envelope
surrounds the B-type star. The double peak structure of the H\,I emission and the presence of V/R variability suggests 
a disc-shaped envelope. V/R variability is usually observed in Be stars and interpreted as oscillations of density enhancements 
and explained by the one-armed oscillation theory \citep{1991PASJ...43...75O}.  Furthermore, quasi-cyclic
photometric variability in time scales of thousands of days, as observed in this system, has been sometimes reported in Be stars. We notice that the observed long-cycle period of 2500 days (6.8 years) is close to the
average of the V/R variability time-scale, viz.\,7 years \citep{1991A&A...241..159M} .
The above suggests a  Be star nature for this system.
The Classical Be stars are rapidly rotating non-supergiants B-type stars that show or have shown Balmer line emission in the past \citep{2013A&ARv..21...69R}. The emission is formed in a circumstellar disk by electron excitation and subsequent cascade 
recombination in the circumstellar material, mostly neutral hydrogen.
On the observational side, the system
show similarities to the Be star ABE-A01, showing outbursts with a quasi-period of 91.23 days \citep{2018AJ....155...53L}.

At maximum we observe two sets of radial velocities clearly separated and different from the radial velocities observed at minimum. 
This fact, along with the shape of the light curve, might indicate that the system is a binary of orbital period 141.45 days with a dense circumstellar disc 
and an eccentric orbit.
Furthermore,  we notice that at maximum the H$\alpha$ and Paschen emission are larger
and the NaD lines are stronger, indicating possibly a mass transfer episode in an eccentric binary system when the unseen secondary star passes at periastron, overflowing its Roche lobe and depositing material onto the Be star disc. Supporting this  accretion scenario, the detection of the infrared Ca\,II triplet in emission in Be stars  has been interpreted in terms of binarity and stream accretion \citep{2012arXiv1205.2259K, 2018A&A...609A.108S}. However, at the present stage, we cannot discard the possibility that during  
periastron passage an increased injection of mass might occur from the surface of the Be star
into the disk, producing the observed brightenings (ejection scenario).

Future challenges include the determination of the nature of the unseen secondary star and the determination of the orbital parameters, including an explanation for the possible eccentricity. The secondary is probably not a compact object, because of the lack of X-ray and high-excitation lines, and also because the outbursts are of low amplitude, reflecting - in the accretion scenario - the fall of material into a shallow gravitational potential. On the other hand, the secondary star should be much less luminous than the Be star to remain undetected in the spectrum. This suggests a lower temperature main-sequence secondary star. In the ejection scenario, its mass should be not so low in order to perturb appreciably the gravitationally bounded material in the surface of the Be star. Another possibility is a low-mass OB subdwarf secondary, as seen in  HR\,2142 \citep{2016ApJ...828...47P}
and other recently found Be + sdO Galactic binaries \citep{2018ApJ...853..156W}. These relatively hot but very faint objects 
were detected only in combined spectra with very high signal to noise obtained in the ultraviolet spectral region by the {\it International Ultraviolet Explorer} satellite. 

\begin{table}
\centering
\tiny
 \caption{Radial velocities of absorption lines observed during minimum. The observed wavelengths
 are not barycentric corrected (the velocities do).}
 \begin{tabular}{@{}lcc@{}}
 \hline
$\lambda$ (lab) &$\lambda$ (obs) &RV  \\ 
(\AA)& (\AA)& (km s$^{-1}$)\\
\hline
H\,I& &\\
3663.406&	3664.943&	125.8\\
3664.679&	3666.252&	128.7\\
3666.097&	3667.666&	128.3\\
3667.684&	3669.212&	124.9\\
3669.466&	3671.035&	128.2\\
3671.478&	3673.019&	125.8\\
3673.761&	3675.316&	126.9\\
3676.365&	3677.916&	126.5\\
3679.355&	3680.980	&132.4\\
3682.810&	3684.333&	124.0\\
3686.833&	3688.414&	128.6\\
3691.557&	3693.054&	121.6\\
3697.154&	3698.736&	128.3\\
3703.855&	3705.404&	125.4\\
3711.973&	3713.678&	137.7\\
3721.940&	3723.517&	127.0\\
3734.370&	3735.981&	129.3\\
3750.154&	3751.652&	119.8\\
3770.632&	3772.090	&115.9         \\
3797.900&	3799.448&	122.2\\
3835.386&	3836.957&	122.8\\
3889.051&	3890.670	&124.8\\
4101.737&	4103.427&	123.5\\
\hline
Fe\,I&&\\
4307.902&	4309.796&	120.9\\	
4404.750&	4406.709&	122.4\\	
\hline
Fe\,II&&\\
4508.288&	4510.250&	119.6\\	
4520.224&	4522.205&	120.5\\	
4522.634&	4524.651&	122.8\\	
4555.893&	4557.738&	110.5\\
4848.235&	4850.341&	119.3\\	
5284.109&	5286.394&	118.7\\
\hline
Ti\,II&&\\
4320.960&	4322.861&	121.0\\	
4443.798&	4445.904&	131.2\\	
4464.450&	4466.479&	        125.3\\	
4501.272&	4503.293&	123.7\\
4533.966&	4536.075&	128.5\\
4911.193&	4913.348&	120.6\\	
4563.761&	4565.828&	124.9\\
4571.969&	4574.048&	125.4\\	
\hline
Mg\,II&&\\
4481.226&	4483.126&	116.2\\	
\hline		
Cr\,II & & \\
4824.127&	4826.172&	116.2\\	
4876.440&	4878.554&	119.1\\	
	\hline
Average $\pm$ std  &- &123.9  $\pm$ 5.0 \\
\hline
\end{tabular}
\end{table}

\begin{table}
\centering
\small
 \caption{Radial velocities of absorption lines observed during maximum. }
 \begin{tabular}{@{}lcc@{}}
  \hline
$\lambda$ (lab) &$\lambda$ (obs) &RV  \\ 
(\AA)& (\AA)& (km s$^{-1}$)\\
\hline
H&&\\
4340.472 &	4342.93	&174.3\\
4101.737	&       4103.94	&165.5\\
3970.075&	3972.31	&173.3\\
3889.051&	3891.03	&157.1\\
3835.386&	3837.63	&179.9\\
\hline
     Fe\,II &&\\
4178.855&	4181.03	&160.5\\
4233.167&	4235.48	&168.3\\
4549.467&	4551.92	&166.1\\
4555.893&	4557.99	&142.5\\
4583.829&	4586.13	&155.0\\
4629.336&	4631.76	&161.5\\
4924.043&	4926.61	&160.8\\
5018.434&	5021.04	&160.2\\
5173.002&	5175.34	&140.0\\
5197.569&	5200.14	&152.8\\
5275.994&	5278.41    &141.9\\
5316.609&	5319.23	&152.3\\
\hline
Ca\,II &&\\
3933.660&	3935.45   &140.9\\
3968.470&	3970.21   &135.9\\
\hline
Mg\,II&&\\
4481.226&	4483.17   &134.1\\
\hline
Ti\,II&&\\
4443.798&	4446.31	&174.0\\
4533.966&	4536.01	&139.6\\
4563.761&	4565.83	&140.4\\
4571.969&	4574.3      &157.3\\
\hline
Cr\,II&&\\
4558.659&	4560.97	&156.5\\
\hline
Fe\,I&&\\
6078.500 &6081.06	        &130.7 \\
6419.980&6717.13	        &156.5\\
\hline
Average & - & 154.0 $\pm$ 13.0 \\
\hline
\end{tabular}
\end{table}

 \section{Conclusions}
 \label{S:6}

Based on the study of 17.5 years of $I$- and $V$-band OGLE photometry and new
X-SHOOTER spectra we find that OGLEJ005039.05-725751.4 is very likely a binary
 consisting of a  Be star in an eccentric orbit with orbital period 141.45 days.
 We also detect a long-cycle of 2500 days in the photometric dataset. 
The strange character of the light curve might be explained by periodic mass transfer in a complex 
binary system as happens in certain types of cataclysmic variables.
 This is supported by our finding that at maximum the H$\alpha$ emission is larger,  the infrared Calcium triplet is seen in emission and the NaD lines are stronger.
 A detailed explanation of the system is beyond the capabilities of the available data. Future plans include
the acquisition of time resolved spectroscopy to resolve the binary orbit and enlighten the brightening and line emission enhancement episode.

\section{Acknowledgments}
\label{S:7}

 Thanks to the anonymous referee for providing useful comments that improved the first version of this manuscript.
This research has made use of the SIMBAD database, operated at CDS, Strasbourg, France.
R.E.M. acknowledges support by VRID-Enlace 218.016.004-1.0,  VRID-Enlace 216.016.002-1.0 and the BASAL Centro de Astrof{\'{i}}sica y Tecnolog{\'{i}}as Afines (CATA) PFB--06/2007. 
LC acknowledges financial support from the Agencia de Promoci\'on Cient{\'{i}}fica y Tecnol\'ogica (Pr\'estamo BID PICT 2016/1971), CONICET (PIP 0177), and the Universidad Nacional de La Plata (Programa de Incentivos G11/137), Argentina.
L.C. thanks also support from the project CONICYT + PAI/Atracci\'on de 
capital humano avanzado del extranjero (folio PAI80160057).
The OGLE project has received funding from the Polish National Science
Centre grant MAESTRO no. 2014/14/A/ST9/00121.












\end{document}